\documentclass[twocolumn,pre,superscriptaddress]{revtex4}
\usepackage{tikz}
\usepackage{graphicx}%
\usepackage{color} 
\usepackage{transparent}
\usepackage{psfrag}
\usepackage{dcolumn}
\usepackage{amsmath,wasysym}
\usepackage{bm} 
\usepackage{amssymb}
\usepackage{latexsym}
\usepackage{hyperref} 
\usepackage{booktabs}
\usepackage{latexsym, bbold}
\begin{document}

\def\bea{\begin{eqnarray}}
\def\eea{\end{eqnarray}}
\def\beq{\begin{equation}}
\def\eeq{\end{equation}}
\def\f{\frac}
\def\k{\kappa}
\def\e{\epsilon}
\def\ve{\varepsilon}
\def\be{\beta}
\def\D{\Delta}
\def\h{\theta}
\def\t{\tau}
\def\a{\alpha}

\def\cDa{{\cal D}[X]}
\def\cD{{\cal D}[x]}
\def\cL{{\cal L}}
\def\cLo{{\cal L}_0}
\def\cLa{{\cal L}_1}
\def\rv{{\bf r}}
\def\tv{\hat t}
\def\on{{\omega_{\rm on}}}
\def\off{{\omega_{\rm off}}}
\def\fv{{\bf{f}}}
\def\fm{\bf{f}_m}
\def\zh{\hat{z}}
\def\yh{\hat{y}}
\def\xh{\hat{x}}
\def\km{k_{m}}

\def\Re{{\rm Re}}
\def\sj{\sum_{j=1}^2}
\def\rk{\rho^{ (k) }}
\def\rek{\rho^{ (1) }}
\def\cek{C^{ (1) }}
\def\rz{\rho^{ (0) }}
\def\rt{\rho^{ (2) }}
\def\rtb{\bar \rho^{ (2) }}
\def\trk{\tilde\rho^{ (k) }}
\def\trek{\tilde\rho^{ (1) }}
\def\trz{\tilde\rho^{ (0) }}
\def\trt{\tilde\rho^{ (2) }}
\def\r{\rho}
\def\tD{\tilde {D}}

\def\s{\sigma}
\def\kb{k_B}
\def\bF{\bar{\cal F}}
\def\F{{\cal F}}
\def\la{\langle}
\def\ra{\rangle}
\def\nn{\nonumber}
\def\up{\uparrow}
\def\dn{\downarrow}
\def\S{\Sigma}
\def\dg{\dagger}
\def\d{\delta}
\def\p{\partial}
\def\l{\lambda}
\def\L{\Lambda}
\def\G{\Gamma}
\def\o{\Omega}
\def\w{\omega}
\def\g{\gamma}
\def\E{{\mathcal E}}

\def\O{\Omega}

\def\vv{ {\bf v}}
\def\jv{ {\bf j}}
\def\jr{ {\bf j}_r}
\def\jd{ {\bf j}_d}
\def\jdd{ { j}_d}
\def\noi{\noindent}
\def\a{\alpha}
\def\d{\delta}
\def\p{\partial} 

\def\la{\langle}
\def\ra{\rangle}
\def\e{\epsilon}
\def\n{\eta}
\def\g{\gamma}
\def\break#1{\pagebreak \vspace*{#1}}
\def\hf{\frac{1}{2}}
\def\rcurs{r_{ij}}

\def\bv{ {\bf b}}
\def\uv{ {\bf u}}
\def\rv{ {\bf r}}
\def\cf{{\mathcal F}}


\title{Semiflexible polymer in a gliding assay: reentrant transition, role of turnover and activity}
\author{Amir Shee}
\email{amir@iopb.res.in}
\affiliation{Institute of Physics, Sachivalaya Marg, Bhubaneswar 751005, India}
\affiliation{Homi Bhaba National Institute, Anushaktigar, Mumbai 400094, India}

\author{Nisha Gupta}
\email{gnisha@iitpkd.ac.in}
\affiliation{Department of Physics, Indian Institute of Technology Palakkad, Palakkad 678557, India}

\author{Abhishek Chaudhuri}
\email{abhishek@iisermohali.ac.in}
\affiliation{Indian Institute of Science Education and Research Mohali, Knowledge City, Sector 81, SAS Nagar 140306, Punjab, India}

\author{Debasish Chaudhuri}
\email{debc@iopb.res.in} 
\thanks{(corresponding author)}
\affiliation{Institute of Physics, Sachivalaya Marg, Bhubaneswar 751005, India}
\affiliation{Homi Bhaba National Institute, Anushaktigar, Mumbai 400094, India}

\date{\today}

\begin{abstract}
We consider a model of an extensible semiflexible filament moving in two dimensions on a motility assay of motor proteins represented explicitly as active harmonic linkers. Their heads bind stochastically to polymer segments within a capture radius, and extend along the filament in a directed fashion before detaching.  Both the extension and  detachment rates are load- dependent and generate an active drive on the filament. The filament undergoes a first order phase transition from open chain to spiral conformations and shows a reentrant behavior in both the active extension and the turnover, defined as the ratio of attachment- detachment rates.  Associated with the phase transition, the size and shape of the polymer changes non-monotonically,  and the relevant autocorrelation functions display double- exponential decay. The corresponding correlation times show a maximum signifying the dominance of spirals. The orientational dynamics captures the rotation of spirals, and its correlation time decays with activity as a power law.
   
\end{abstract}

\maketitle

\section{Introduction}
The cytoskeleton in living cells consists of semiflexible filaments like F-actins and microtubules, 
and motor proteins\,(MP)~\cite{Alberts2009, Howard2005}. 
The MPs hydrolyse ATP to undergo binding, unbinding cycles and move in a directional manner along the associated filaments~\cite{Julicher1997b, Vale2003, Chowdhury2013}. 
 On cross-linked filaments of cytoskeleton, the  active chemical cycle of MPs generate mechanical stress to maintain the cell structure and dynamics~\cite{Fletcher2010,Julicher2018a}. 
 The MPs drive energy flux at the smallest length scales of the system, typical of active matter~\cite{Needleman2017, Vicsek2012, Marchetti2013}.
 This breaks the detailed balance, and the equilibrium fluctuation- dissipation relation.

The {\em in vitro} molecular motor assays are often used to derive direct physical understanding of the active properties of filaments and MPs~\cite{Kron1986, Howard1989, Vale1994, AMOS1991}.
The motility assay setup with actin filaments or microtubules floating on top of an immobilized MP- bed, showed fascinating dynamical behaviors, e.g., spiral formation, collective gliding and  swirling~\cite{AMOS1991, Bourdieu1995c, Sekimoto1995, Sumino2012, Schaller2010}.
For spiral formation of microtubules on kinesin assay~\cite{AMOS1991, Bourdieu1995c}, a microtubule- specific theory has been recently developed~\cite{Ziebert2015}. However, similar behavior has been observed in other active polymer studies~\cite{Jiang2014, Isele-Holder2015, Eisenstecken2016, Winkler2017, Man2019, Winkler2020}.

In this paper we consider a detailed theoretical model of a two- dimensional motility assay, and study the change in shape and size of an extensible semiflexible polymer driven by MPs. In our model, the MPs are immobilized by attachment of their {\em tails} to a substrate, while the {\em head} domains undergo active attachment- detachment with the filament, and drive the filament by performing active extension. The detachment and extension rates are assumed to be load dependent in a manner consistent with established MP models~\cite{Schnitzer2000,Grill2005}. 
Most of the current studies which attempt to understand the static and dynamic properties of a filament in the presence of activity, either consider the polymers as made up active monomers with a constant velocity in the tangential direction or introduce activity via an active noise term~\cite{ Jiang2014, Chelakkot2014, Ghosh2014,Isele-Holder2015, Shin2015c, Eisenstecken2016, DeCanio2017, Winkler2017, Duman2018, Prathyusha2018, Man2019, Mokhtari2019, Peterson2020, Anand2020, Winkler2020}. 
However, two-fold effect of MPs on the conformational and dynamical properties of a semiflexible filament are profound and therefore need explicit consideration~\cite{Chaudhuri2016c, Gupta2019, Foglino2019}.

We perform extensive numerical simulations to study the polymer in motility assay, and use phenomenological arguments to illustrate several findings.  
We obtain a {\em first order} conformational transition from open chain to spiral as a function of the  MP activity, which has two main aspects: (i)~the rate of extension, and (ii)~the turnover -- given by the ratio of attachment- detachment rates. The transition is characterized by the coexistence of the open and spiral phases. Obtaining the resultant phase diagram is the { first main contribution of this paper}.  It shows a remarkable reentrance from open chain to spiral to open chain with increasing activity.  The spirals are characterized by their turning number. { An approximate  data- collapse of the non- monotonic variations of the mean squared turning number with active extension for different turnovers leads to a scaling function. This is supported by a torque- balance argument, which also describes the phase boundary.} {This is our second main contribution}.  
The distribution function of the end- to- end separation shows bi-stability capturing the coexistence between open and spiral states. We use radius of gyration tensor to determine the instantaneous size, shape, and effective orientation of the polymer. Accompanied by the reentrance transition, the polymer size, and shape- asymmetry show non-monotonic variations with activity. { The non- monotonic variation in size shows qualitative  difference with respect to that of polymers in active bath~\cite{Eisenstecken2016}.}  We study the steady state dynamics using the two- time autocorrelation functions.  The dynamics of turning number, size and shape of the polymer depend  on the conformational changes. Their autocorrelations reveal double- exponential decay at phase- coexistence, corresponding to the relaxation within a state, and slow transition between the states. The correlation time shows non-monotonic variation with a maximum at an intermediate rate of MP extension. { This is our third main result.} The autocorrelation function of the instantaneous orientation of the polymer conformation shows an overall single time- scale decay, and oscillations related to the rotation of the spirals at higher activity. The corresponding correlation time decreases with MP extension rate as a power- law.

The plan of the paper is as follows. 
In Sec.\ref{sec:model} we present the detailed model of the motility assay and the extensible semiflexible polymer. 
We present our results in Sec.~\ref{sec_results}. 
In Sec.~\ref{sec_spiral} we demonstrate the spiral formation with the help of turning number. Using its probability distribution, in Sec.~\ref{sec_pt}, we demonstrate a first order phase transition from open chain to spiral with increasing activity. 
The phase diagram is presented in Sec.~\ref{sec_pd}.
In Sec~\ref{sec_turn} we discuss an approximate scaling form of the turning number fluctuations. 
In Sec.~\ref{sec_sz_shp} the end- to- end distribution function,  the change in polymers size, and shape is discussed. This is followed by a discussion of the polymer dynamics in terms of autocorrelation functions of turning number, polymer size, shape, and orientation in Sec.~\ref{sec_dyn}. Finally, we conclude in Sec.~\ref{sec_dis} summarizing our main results. 


\section{Model and simulation}
\label{sec:model}
We consider an extensible semi-flexible  polymer of $N$-beads with monomer positions $\rv_1,\,\rv_2,\,\dots,\rv_N$. 
The chain is described by both stretching and bending energy terms. 
The bond vectors $\bv_i = \rv_{i+1} - \rv_i$  are defined for $i=1,2,\dots,N-1$ and are oriented along the local tangents $\tv_i = \bv_i / |\bv_i |$. 
The connectivity of the chain is maintained by the stretching- energy 
\bea
\E_s = \sum_{i=1}^{N-1} \f{A}{2 r_0} \left[ \bv_i - { r_0} \tv_i  \right]^2,
\eea 
characterized by the bond- stiffness $A$ and the equilibrium bond- length $r_0$. 
The bending rigidity $\k$ of the semiflexible filament leads to a bending energy cost between  the consecutive tangent vectors, 
\bea
\E_b = \sum_{i=1}^{N-2} \f{\k}{2 r_0} \left[ \tv_{i+1} - \tv_i  \right]^2.
\eea
The self-avoidance of the filament is implemented through a short-ranged Weeks-Chandler-Anderson repulsion between all the non-bonded pairs of beads $i$ and $j$, 
\bea
\E_{\rm WCA} &=& 4[(\s/\rcurs)^{12} - (\s/\rcurs)^6 + 1/4]~{\rm if}~ \rcurs < 2^{1/6}\s \nn\\
&=& 0, ~{\rm otherwise.}
\eea 
Thus the full polymer model is described by the energy cost $\E = \E_s + \E_b + \E_{\rm WCA}$. 
The energy and length scales are set by $\e$ and $\s$ respectively. The corresponding microscopic time scale is $\t_0 = \s\sqrt{m/\e}$. 

\begin{figure}[!t]
	\centering
	\includegraphics[width=7cm]{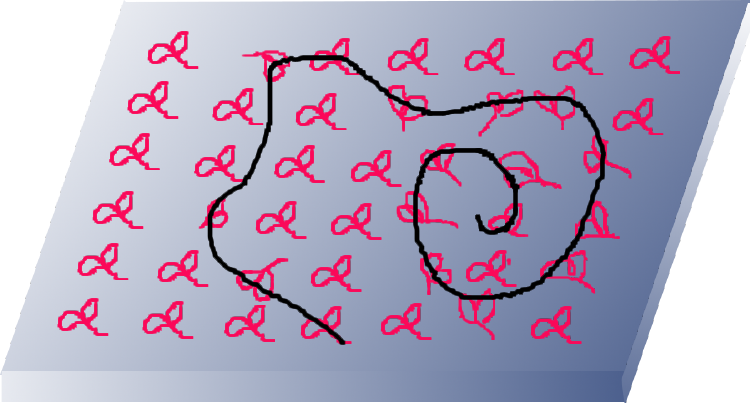}
		\caption{(color online) A schematic diagram of the system showing a polymer floating on the motility assay. The {\em tails} of MPs are attached irreversibly on a square grid. The {\em head} domains can attach to the filament, when any segment of it comes within the capture radius. The active extension of the attached {\em head} along the filament generates force in the  opposite direction. The MP {\em stalk} is modeled as a harmonic spring. }
	\label{figschem}
\end{figure}

In the motility assay setup, the polymer is placed on a substrate of MPs (Fig.~\ref{figschem}\,).  The MPs are assumed to be immobilized by attachment of their {\em tails} irreversibly to the substrate at positions $\rv_0^i=(x_0^i,y_0^i)$ placed on a two dimensional square lattice with a uniform density $\r$.  
The {\em heads} of MPs can attach to the nearest bead of the polymer within a capture radius $r_c$ through a diffusion limited process which is implemented by a constant attachment-rate $\on$. The stalks of the MPs are modeled as elastic linkers of stiffness $k_m$. The extension $\Delta \rv$ of a MP in the attached state generates an elastic force $\fv_l = -k_m\D\rv$ on the segment of the filament it is attached to. 
This extension can be due to two processes: (i)~the attached head may be dragged by the filament, and (ii)~it can move actively over the filament towards one of its ends.
{
The attached head moves along the bonds. Thus its instantaneous location can be anywhere between the beads. The MP- extension generates a force $\fv_l$, which is divided between the beads forming the bond on which the MP- head is located. This is done using the lever rule, and depends on the relative separation of the MP- head with respect to the polymer beads. The nearer the MP is to a specific bead, the larger is the share of the force on it. }
For example, attached kinesins (dyeneins) move along the microtubule towards its positive (negative) end. The active velocity is known to decrease with resistive load, and can be modeled as~\cite{Schnitzer2000, Chaudhuri2016c}
\bea
v^a_t(f_t) = \f{v_0}{1+d_0\exp(f_t/f_s)},
\label{velc} 
\eea
where $f_t=-\fv_l.\tv$, $d_0 = 0.01$ and $f_s$ is the stall force. Here $v_0$ denotes the velocity of free MP. The actual extension $\D\rv$, and as a result $f_t$ on different MP is different. It depends on the time spent in the attached state, which in turn depends on the stochastic detachment rate 
 \bea 
\off = \w_0\exp(f_l/f_d),
\label{eq:off}
\eea 
where $\w_0$ is the bare off rate, $f_l = |\fv_l|$ and $f_d$ sets the scale of the detachment force. The ratio $\on:\off$ does not obey detailed balance. The net force imparted by MPs depend on the  processivity $\o(f_l) = \on/(\on+\w_0\exp(f_l/f_d)\,)$.

We perform molecular dynamics simulations of the polymer using beads of unit mass $m=1$, in the presence of a Langevin heat bath of isotropic friction per bead $\g = 1/\t_0$ keeping the temperature constant at $\kb T=1.0\,\e$. 
We use bond- stiffness $A = 100\,\e/\s$ for the $N=64$ bead chain. 
In equilibrium worm-like-chain, the ratio of the contour length $L=(N-1) r_0$  to persistence length $\l= 2 \k/[(d-1) \kb T]$, the rigidity parameter $u=L/\l$, determines whether the filament behaves like a rigid rod or a flexible polymer~\cite{Dhar2002,Chaudhuri2007}. { The end-to-end distribution of worm-like-chain shows Gaussian chain behavior with a single maximum at zero- separation at $u \approx 10$, and a rigid-rod behavior with a single peak near full extension of the chain at $u \approx 1$.  In the semiflexible regime of $u=$3 to 4, the free energy shows a characteristic double minimum corresponding to the coexistence of both the rigid rod and flexible chain behaviors.} To probe this regime, we choose $\k/r_0  \kb T = 9.46$ corresponding to $u = 3.33$.  
{Unless stated otherwise, we choose the equilibrium bond-length $r_0 = 1.0\,\s$.}
At this point it is important to note that the typical size of individual MPs are three to four orders of magnitude smaller than the typical length of polymers used in motility assay setups. Incorporating this large length scale separation makes the numerical simulations prohibitively expensive. We use a capture radius $r_c=0.5\,\s$, and MP density $\rho = 3.8\, \s^{-2}$ in our simulations. To avoid introduction of further energy scales, we use $k_m = A/\s$. 
To maintain active forces larger than thermal fluctuations, 
we use $f_s = 2\, \kb T/\s$, $f_d = f_s$.  

{
The dynamics of the active system is characterized by the dimensionless ratio of attachment and detachment rates $\on/\w_0$, and a dimensionless P{\'e}clet number $Pe = v_0 L/D_t$ expressed as a ratio of convective and diffusive transport of the filament. Using translational  diffusion coefficient of polymer $D_t = D/(L/r_0)$ 
with $D=\kb T/\g$, one obtains $Pe= v_0 L^2/D r_0$. 
} 
{
This expression, along with the rigidity parameter $u=L/\l$, give the flexure number $Pe\, u = v_0 L^3 (d-1)/2\k r_0$, which plays crucial role in determining buckling instability, and spiral formation in active polymers~\cite{Sekimoto1995, Chelakkot2014, Isele-Holder2015}. The characteristic time for the filament to diffuse over its contour length $L$ is $\t = L^3 \g/4 r_0 \kb T$. We use this as a unit of time in expressing the time-scales in simulation results. 
The numerical integrations are performed using 
$\d t$ adjusted for numerical stability. The presence of turnover reduces the effective active force imparted on the chain, as MPs detach under longer extension. As a result, the smallest $\d t$ required in these simulations is $1.6 \times 10^{-8}\,\t$,} 
{
larger than that was necessary for active polymer simulations~\cite{Isele-Holder2015}.}
The results are presented here from simulations over $2 \times 10^9$ steps, discarding the first $10^9$ steps to ensure steady state measurements. 
\begin{figure}[!t]
	\centering
		\includegraphics[width=8cm]{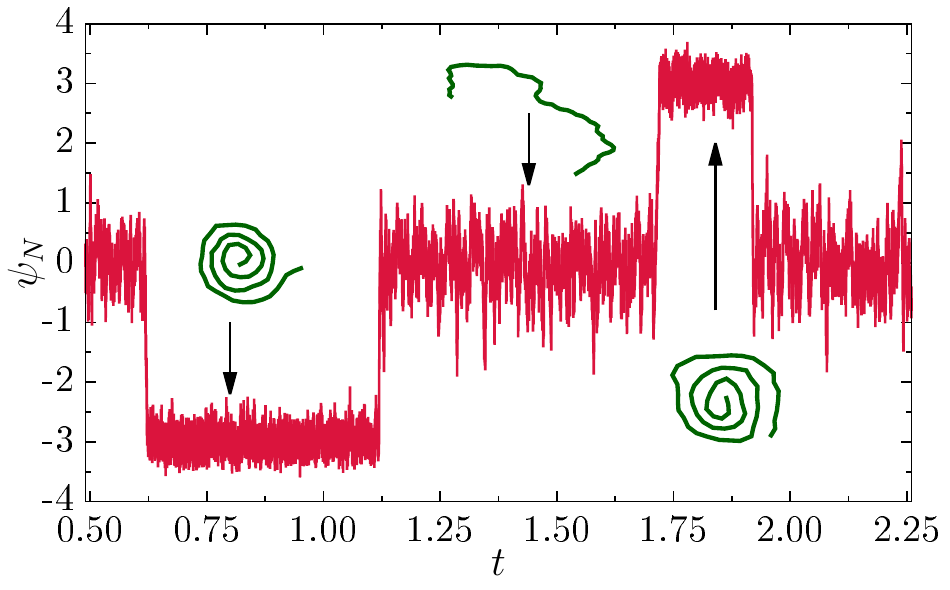} 
	\caption{(color online) Time evolution of the turning number $\psi_N$ at $Pe=10^{5}$ and the ratio $\on/\w_0=1$. Time $t$ is expressed in the unit of $\t$. The plot shows stochastic switching between three states, an open state with $\psi_N \approx 0$,  and two spiral states with $\psi_N \approx \pm 3$. Representative polymer configurations corresponding to the three states are shown at three time instances indicated by arrows.}
	\label{fig:psi_vs_t}
\end{figure}

 \begin{figure}[t]
\centering
\includegraphics[width=8.6cm]{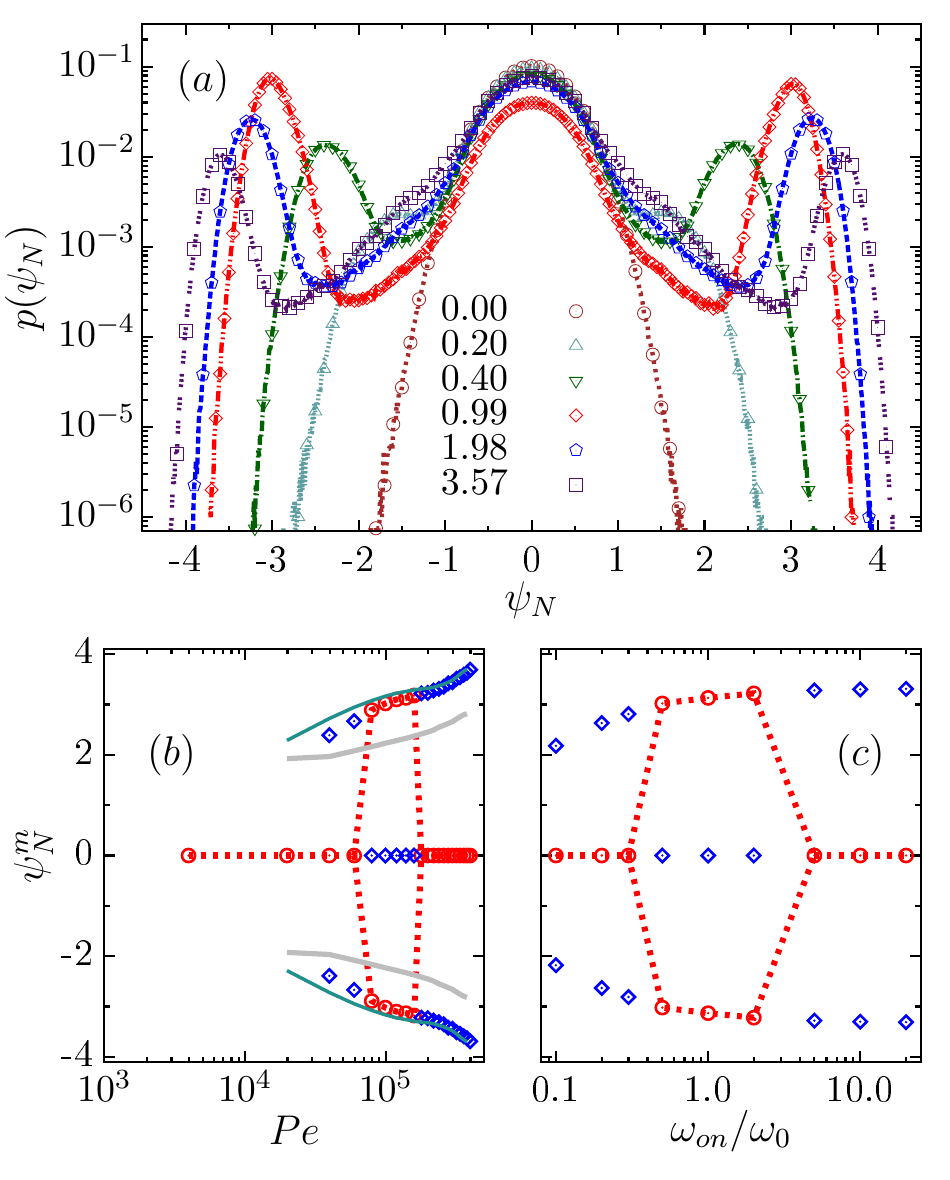}
\caption{(color online) $(a)$ Probability distribution functions of turn number $p(\psi_{N})$ for different $Pe= \tilde{Pe} \times 10^5$ where values of $\tilde{Pe}$ are denoted in the figure legend,  
at a fixed ratio  $\omega_{on}/\omega_{0}=1$. The triple- maxima characterize the coexistence in transition from open chains to spirals. The dependence of the stable (global) (red $\Circle$) and metastable (blue $\Diamond$) maxima of $p(\psi_{N})$ are shown as a function of $Pe$ at $\on/\w_0 =1$ in ($b$), and as a function of $\on/\w_0$ at $Pe=1.39\times 10^5$ in ($c$). In ($b$), the green lines show the plot of $\pm (|u_4|/2u_6)^{1/2}$, and the grey lines show the plot of $\pm(u_2/2 u_4)^{1/2}$, where $u_2$, $u_4$ and $u_6$ are defined by Eq.\eqref{eq_fe}.  
}
\label{fig:psiL}
\end{figure}

\section{Results}
\label{sec_results}
In this section we present the results of numerical simulations. 

\subsection{Formation of spiral}
\label{sec_spiral}
Beyond a minimum on- off ratio $\on/\w_0$, and activity $Pe$, the polymer  spontaneously starts to get into spiral structures.  The nature of the spiral can be quantified in terms of the turning number~\cite{Krantz1999}, $\psi_i = (1/2\pi)  \sum_{j=1}^{i-1} [\theta_{j+1} - \theta_j]$ where $\h_j$ is defined by $\tv_j = (\cos \h_j, \sin \h_j)$, and
$[\theta_{j+1} - \theta_j]$ gives the angle increment between consecutive bonds. 
Thus turning number $\psi_N$ with $i=N$ measures the (real) number of turns the chain takes between its two ends.  
For a straight chain $\psi_N=0$, and for a chain forming a single anticlockwise (clockwise) loop $\psi_N=1$\,($\psi_N=-1$). 
Larger values of $\psi_N$ correspond to more than one turn forming the spiral. 
In Fig.\ref{fig:psi_vs_t}) we show a typical time series of $\psi_N$, along with three  representative conformations corresponding to $\psi_N \approx 0$ and a turning number $\psi_N \approx \pm 3$.    

\subsubsection{First order phase transition: open chain to spiral}
\label{sec_pt}
In Fig.~\ref{fig:psiL}($a$) we show the steady state probability distributions of the turn number 
$p(\psi_N)$ at different values of $Pe$, corresponding to a fixed on- off ratio $\on/\w_0 = 1$. At small $Pe$ we find a unimodal distribution with the maximum located at $\psi_N=0$ corresponding to open chains. With increasing the activity to $Pe = 0.2 \times 10^5$ two other metastable maxima appear in $p(\psi_N)$  near $\psi_N=\pm 1.8$, positioned symmetrically around the central peak at $\psi_N=0$, which remains the global maximum. 
Appearance of such metastable states across a phase transition is a characteristic of a first order transition.  As we increase $Pe$, the heights of the maxima corresponding to spiral grow.
Near { $Pe = 0.67 \times 10^5$}, all the three maxima of $p(\psi_N)$ becomes equally probable, identifying  the {\em binodal point} of the first order phase transition from the open- chain to spiral (data shown in ESI~\footnote{see DOI: 10.1039/D0SM01181A}). 
The increase in the probability of the spiral states, characterized by the rise of height of the two non-zero $\psi_N$ maxima, continue up to { $Pe = 1.19 \times 10^5$}. This indicates further (de-)~stabilization of the (open)~spiral state.
A remarkable non-monotonic feature is observed with further increase in $Pe$.   
For larger $Pe$, the non-zero $\psi_N$- peaks corresponding to the spiral states start to reduce in height with respect to the peak at $\psi_N=0$. Again near $Pe=1.58 \times 10^5$, all three maxima attain the same height, indicating a {\em binodal} corresponding to the {\em reentrant transition} back from spiral to open chain state. At larger $Pe$, the heights of the non-zero $\psi_N$ peaks keep diminishing with increasing $Pe$ values. 
Despite this non-monotonic nature of the stability of open and spiral states, it should be noted that, all through, the positions of the peaks at non-zero turning number $\psi_N$ consistently increases  to larger amplitudes of $\pm \psi_N$ as $Pe$ increases. Thus, while the probability of spirals at $Pe > 1.58 \times 10^5$ gets smaller with increasing $Pe$, when formed,  the spirals at higher $Pe$ consistently display higher turning numbers. 

We analyzed all such probability distributions within a range of $0 \leq Pe \leq 3.97 \times 10^5$, and attachment- detachment ratios $0.1 \leq \on/\w_0 \leq 20$ using the locations and heights of the peak positions of $p(\psi_N)$. In Fig.~\ref{fig:psiL}($b$), we show using $\Circle$ (red), the $\psi^m_N$ values corresponding to the stable phase, i.e., the peak position(s) of the global maximum (maxima) in $p(\psi_N)$. Points denoted by $\Diamond$ (blue) show the peak positions corresponding to the metastable state(s), having peak heights smaller than the global maximum. The dotted lines are guide to eye showing the variation of the global maximum with increasing $Pe$, which displays the open- to predominantly spiral- to  predominantly open transition as expected from the probability distributions. Note that the coexistence points, symmetric about the central peak (points corresponding to $\psi^m_N = 0$) mark the familiar coexistence curves (binodal) in a first order phase transition. The various transitions are the unique non-equilibrium features of the motility assay set up. Similar non-equilibrium features are observed when $\psi_N^m$s is plotted as a function of the ratio $\on/\w_0$ at a constant $Pe$ (Fig.~\ref{fig:psiL}($c$)).  

From the probability distribution of $\psi_N$, and using an effective {\em equilibrium}- like approximation $p(\psi_N) \sim \exp[-\cf(\psi_N)]$ 
we can write 
\bea
\cf(\psi_N) = \hf u_2 \psi_N^2 - u_4 \psi_N^4 + u_6 \psi_N^6,
\label{eq_fe}
\eea
apart from an additive constant.  Such a fitting with Fig.~\ref{fig:psiL}($a$) allows us to obtain 
the values of $u_2$, $u_4$ and $u_6$ as a function of $Pe$ and $\on/\w_0$ ratio.  
It is straightforward to show~\cite{Chaikin1995} that along the first order line { described by $\p \cf/\p \psi_N =0$ and $\cf =0$}, the turn number obeys the relation $\psi_N = \pm (|u_4|/2u_6)^{1/2}$. This shows good agreement with simulation results~(Fig.~\ref{fig:psiL}($b$)\,).  Moreover, the spinodal lines, { obeying $\p \cf/\p \psi_N =0$ and $\p^2 \cf/\p \psi_N^2 =0$,}  are given by $\psi_N = \pm(u_2/2 u_4)^{1/2}$, and are shown by the grey lines in Fig.~\ref{fig:psiL}($b$).

\begin{figure}[!t]
	\centering
	\includegraphics[width=8cm]{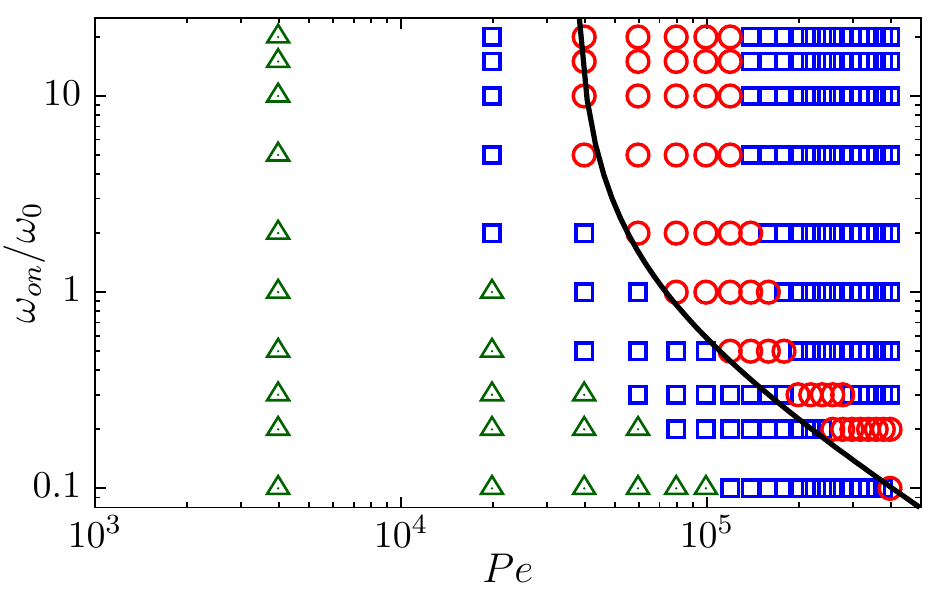} 
	\caption{(color online) Phase diagram calculated from probability distribution of turn number $p(\psi_N)$. The data points denoted by green $\triangle$ denotes a stable open chain state, in the  complete absence of spirals. The blue $\Box$ points denote stable open chains in the presence of metastable spirals. The red $\Circle$ denotes stable spirals coexisting with metastable open chains. The boundaries between $\Box$ and  $\Circle$ denote the binodals where open chains and spirals are equally probable. The solid line capturing one such phase boundary is a plot of the function $\on/\w_0 = \a/(Pe-\a)$ where $\a=3.67 \times 10^4$~(see Sec.\ref{sec_turn}).} 
	\label{fig:phdia}
\end{figure}

\begin{figure*}[!t]
	\centering
	\includegraphics[width=16cm]{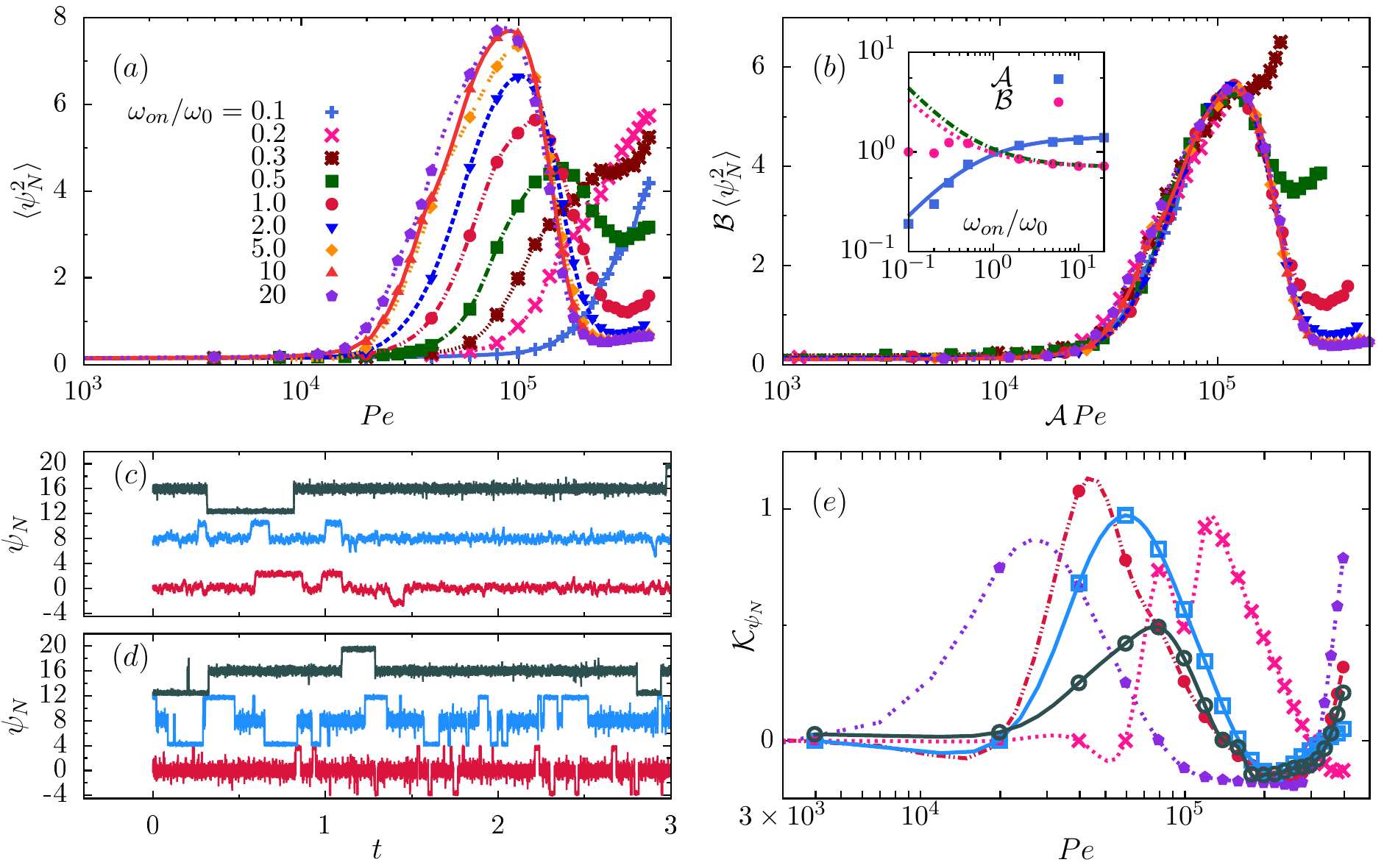}
	\caption{(color online) $(a)$~The steady state turning number fluctuation $\la \psi_N^2 \ra$ as a function of $Pe$ at different values of $\on/\w_0$ ratios denoted in the figure legend. ($b$)~Approximate data collapse of different plots in ($a$) by using  scale factors ${\cal A}$, ${\cal B}$. Inset: The dependence of ${\cal A}$ and ${\cal B}$ on $q=\on/\w_0$. The solid line $\o_f=1.42\, q/(0.52 + q)$ shows a  fit to the data for ${\cal A}$.  The dotted line shows a fit $(0.37+q)/1.41\,q$ to the data for ${\cal B}$ in the regime $q > 0.3$. The dash- dotted line is a plot of $1/\o_f$. Comparison of time series of $\psi_N$ for bond lengths $r_0=1.0\s$~(red), $0.75\s$~(blue) and $0.5\s$~(green) are shown at $Pe=3.97 \times 10^4$~($c$) and $Pe=3.97 \times 10^5$~($d$). Time $t$ is expressed in the unit of $\t$. The data for $r_0=0.75\,\s$ and $0.5\,\s$ are shifted upwards by $8$ and $16$ for better visibility. ($e$) Kurtosis ${\cal K}_{\psi_N}$ of turning number $\psi_N$ plotted at $\on/\w_0=0.2,\,1.0,\,20$ with data shown by the same symbols as in ($a$). The open $\Box$ and $\Circle$ denote data at $\on/\w_0=1.0$ for $r_0=0.75\,\s$ and $0.5\,\s$ respectively.}
	\label{fig:turn}
\end{figure*}

\subsubsection{Phase diagram}
\label{sec_pd}
In Fig.~\ref{fig:phdia}, we plot the phase diagram in the $\on/\w_0$ - $Pe$ plane characterizing the {\em open} and {\em spiral} states and their stability.
The symbol $\bigtriangleup$ denotes the region where the {\em open} chain is the only phase possible, with the distribution $p(\psi_N)$ showing a single maximum at $\psi_N=0$. 
The regions denoted by $\Square$ indicates  coexistence of the stable {\em open chain}, with a metastable {\em spiral} phase. In the region denoted by $\Circle$ in the phase diagram, it is the {\em spiral} state which is stable, but in coexistence with a metastable {\em open} state. The two  boundaries  between the $\Circle$ and $\Box$ in Fig.~\ref{fig:phdia} identify the two {\em binodal} lines of the first- order transition. Along them, both the open and the spiral states are equally probable. The presence of these two binodal lines characterize the reentrant nature of the first order conformational transition of the polymer. 

The phase diagram clearly brings out the importance of the attachment-detachment kinetics of the MPs. At a fixed  $\on/\w_0$, the polymer goes from a stable open chain to stable spiral to stable open chain reentrant transition with increase in $Pe$. At higher attachment-detachment ratios, $\on/\w_0 > 2$ for the parameters in our simulations, the region of stable spiral states $0.4 \times 10^5 \leq Pe \leq 1.19\times 10^5$, is independent of the $\on/\w_0$ ratio. At lower $\on/\w_0$, the region where the spiral state is the stable state of the polymer appears at progressively larger $Pe$ values. Also the single maxima region, corresponding to only an open chain conformation, persists for higher values of $Pe$ at low  $\on/\w_0$.  

{Active polymers showed formation of spirals at activity larger than the onset of buckling instability~\cite{Isele-Holder2015}. However, this did not show the re-entrance behavior we find.  Our detailed modeling of the MP- bed allowed us to clearly characterize the impact of the MP turnover, revealing the dependence on the $\on/\w_0$ ratio. This remained outside the scope of the active polymer model. }

\subsubsection{Turn number fluctuations}
\label{sec_turn}
In this section we consider the first two moments of the $p(\psi_N)$ distributions. This is due to the fact that, with respect to the full distributions, moments are easier quantities to determine from experiments. 
The chiral symmetry in the system $p(-\psi_N) = p(\psi_N)$ ensures that, all through, $\la \psi_N\ra=0$. 
The quantitative measure of the effective turn number is given by the root- mean square fluctuation $\la \psi^2_N \ra^{1/2}$. Fig.~\ref{fig:turn}($a$) shows the non-monotonic variation of  $\la \psi^2_N \ra$ with $Pe$ at fixed $\on/\w_0$ ratios, corresponding to the reentrant transition. 
Re-scaling of $Pe$ and $\la \psi^2_N \ra$ leads to an approximate data collapse as shown in Fig.~\ref{fig:turn}($b$). We can extract a functional dependence of the scale factors ${\cal A},\, \cal{B}$ on the ratio $q =\on/\w_0$ as ${\cal A} \approx \o_f(q)$ and ${\cal B} \approx 1/\o_f(q)$ (see inset of Fig.~\ref{fig:turn}($b$)). $\o_f(q)$ has the form of a bare processivity, $\o(f_l=0) = \on/(\on + \w_0) = q/(1 + q)$. The data- collapse suggests a functional dependence 
\bea
\la \psi_N^2 (q,Pe) \ra \approx \o_f(q) {\cal G}[\o_f(q) \,Pe].
\label{eq_collapse}
\eea

A spiral with radius $R$ has a turn number $\psi_N = L/2\pi R$. The shape can be maintained via a torque balance $F R^2 = \k/R$,  where $F$ denotes the MP force per unit length. 
This force depends $f_l$, the force exerted due to active extension of MPs, the linear density of MPs $\sqrt{\r}$, and their processivity $\O(f_l)$. { The mean of the active force $f_l$ is denoted here by $f_a \approx \g v_0$.} Thus the net active force per unit length $F:= \sqrt{\r} f_a \O(f_a)$. 
This leads to the following activity dependence of turning number 
\bea
\psi_N^2 \sim  {\cal G}_1(\O(f_a), f_a). 
\label{eq_scale}
\eea
Noting that $Pe \sim f_a$, Eq.\eqref{eq_scale} is related to but cannot fully capture the scaling form in Eq.\eqref{eq_collapse}. The reason lies in the fact that the polymer switches between the spiral and open state, and $\la \psi_N^2 \ra$ is averaged over the probability distribution spanning both the states. 

The onset of spiral requires $\psi^2_N > 1$, i.e., $F > F_c= \k (2\pi/L)^3$. 
Thus the phase boundary denoting this is given by $F:=f_a \O(f_a) = F_c$.  
In the limit of load- independent detachment rate, { with $f_a \sim Pe$, the equality $f_a \O(f_a) = F_c$ leads to a dimensionless form $q Pe/(1+q) = \a$, where $q=\on/\w_0$, and $\a$ denotes a dimensionless constant proportional to $F_c$.} This can be simplified to the hyperbolic relation 
\bea
\on/\w_0 = {\a}/(Pe- {\a}).
\label{eq_phbound}
\eea
 In the phase diagram Fig.~\ref{fig:phdia}, the solid line is a plot of this function with $\a=3.67 \times 10^4$,  and  approximately captures the phase boundary of the onset of spiral phase.

As it has been pointed out earlier~\cite{Isele-Holder2015,Duman2018}, the modulation of potential energy along the chain due to WCA repulsion from polymer beads costs energy to slide chain segments past each other. The resultant increase in sliding friction can increase the lifetime of spirals. 
To examine this we consider chains of the same contour length $L$ but smaller bond lengths $r_0=0.75\,\s$ and $0.5\,\s$ having smoother potential profiles along the chain.  In Fig.~\ref{fig:turn}($c$) and ($d$) we show a comparison between their time series of turning number $\psi_N$
at two activities, $Pe=3.97 \times 10^4$, $3.97 \times 10^5$, keeping $\on/\w_0=1$. 
We find formation of spirals in all the cases. As expected, the life-time of spirals corresponding to all the different phases decreases with reduction of $r_0/\s$, smoothening the polymer. For each $r_0/\s$, however, the time-scale shows non-monotonic variation with $Pe$~(Sec.\ref{sec_ts}). 
A quantitative analysis of the time-scales are presented in Fig.~\ref{fig_corrt} of Sec.~\ref{sec_ts} and Appendix-\ref{sec_tc_smooth}.

In Fig.~\ref{fig:turn}($e$) we show variation of  kurtosis  ${\cal K}_{\psi_N} = [\la \psi_N^4 \ra/3 \la \psi_N^2\ra^2 -1]$ with $Pe$ for three different $r_0/\s$ ratios calculated at $\on/\w_0=1$. It also shows ${\cal K}_{\psi_N} (Pe)$ for $\on/\w_0=0.2,\,20.0$ using the chain with bond length $r_0=1.0\,\s$. To reduce statistical uncertainties we calculated kurtosis over several initial conditions such that the distribution of $\psi_N$ gets symmetric and restricting analysis to the spiral states~\cite{Isele-Holder2015}. At small $Pe$ we find ${\cal K}_{\psi_N} =0$, consistent with the Gaussian distribution. As the spirals start to appear $\la \psi_N^4 \ra$ increases, increasing ${\cal K}_{\psi_N}$. At higher $Pe$, as the spirals stabilize, the second cumulant $\la \psi_N^2\ra$ start to dominate reducing ${\cal K}_{\psi_N}$ from a maximum to eventually ${\cal K}_{\psi_N}$ reach a minimum. Finally at further higher $Pe$, the kurtosis increases again corresponding to the re-entrance.

{
The kurtosis ${\cal K}_{\psi_N}$ calculated for three different values of $r_0/\s=1.0,\, 0.75,\, 0.5$ at  $\on/\w_0=1$ in Fig.~\ref{fig:turn}($e$) display similar non-monotonic behavior, but the peak of the curves shift towards larger $Pe$ for smaller $r_0/\s$. For example, the peak position  of ${\cal K}_{\psi_N}$ shifts from $Pe=4.4 \times 10^4$ at $r_0/\s=1.0$ to $Pe=6.0 \times 10^4$ at $r_0/\s=0.75$ and to  $Pe=7.9 \times 10^4$ at $r_0/\s=0.5$.
Such a shift can be understood by noticing that the reduction in $r_0/\s$ increases the bending rigidity $\k$ of the filament.  In Appendix-\ref{sec_lp} and Fig.~\ref{eq_lp} we show how the persistence length of the equilibrium polymer increases with reduction in $r_0/\s$. As has been pointed out before Eq.(\ref{eq_phbound}), the active force $F_c$ needed for the onset of spiral increases linearly with $\k$. Thus spiral formation at smaller $r_0/\s$ requires higher $Pe$.   
\footnote{The availability of more attachment points of MPs for a filament with smaller $r_0/\s$, within our model, could increase the imparted active force on the filament. However, this effect would shift the ${\cal K}_{\psi_N}$ graphs to smaller $Pe$, unlike what we observe in Fig.~\ref{fig:turn}($e$).} 
 
}

\begin{figure}[t]
	\centering
	\includegraphics[width=8cm]{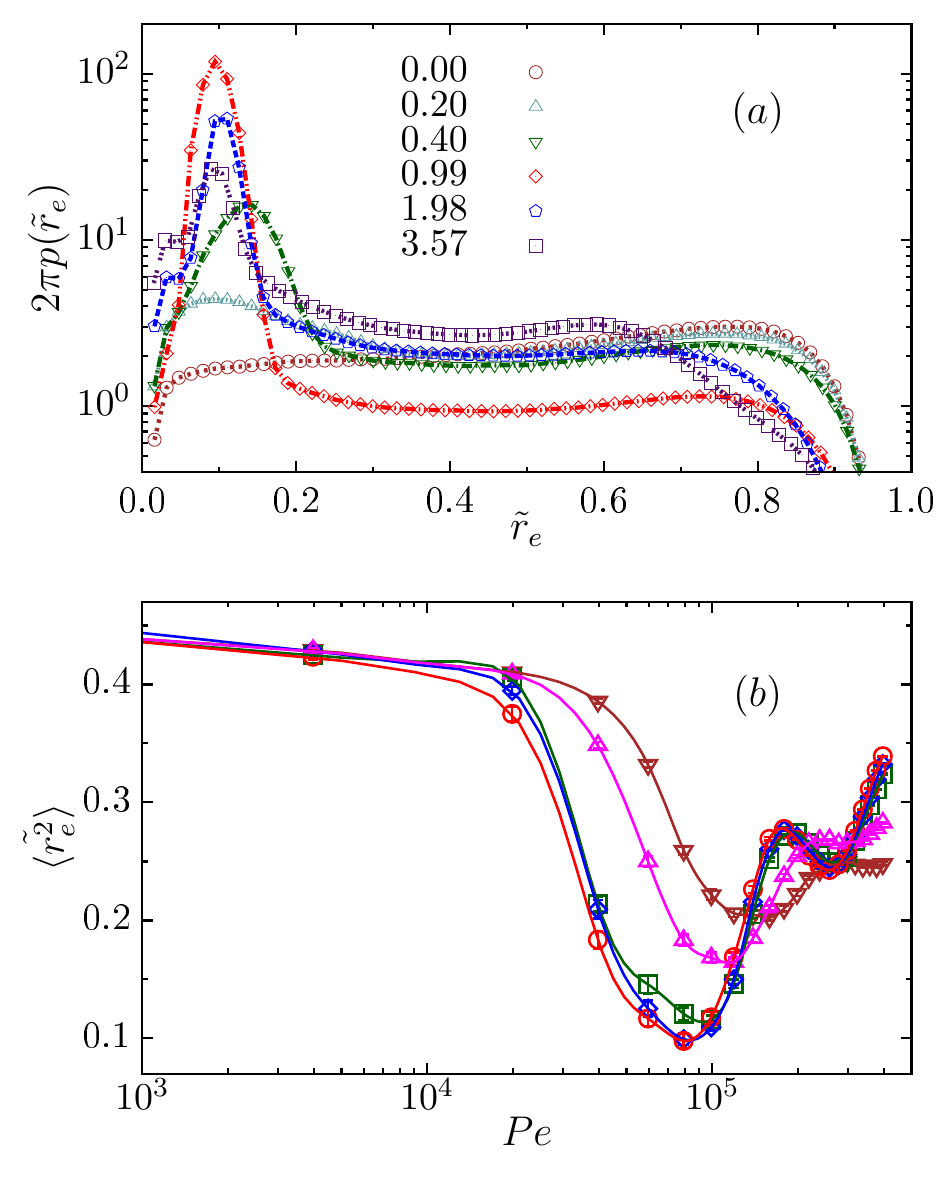} 
	\caption{(color online) $(a)$~End-to-end distribution functions $2\pi p(\tilde{r}_{e})$ for $Pe= \tilde{Pe} \times 10^5$ with $\tilde{Pe}$ values shown in the figure legend, 
	at a fixed $\on/\w_0=1$. 
	$(b)$~Mean squared end- to- end separation $\langle \tilde r_e^2 \rangle$  
	as a function of $Pe$ for $\omega_{on}/\omega_{0}=0.5$\,($\triangledown$), $1$\,($\triangle$), $5$\,($\Box$), $10$\,($\Diamond$), $20$\,($\Circle$). Error bars are smaller than the symbol size. The lines through data are guides to eye. }
	\label{fig_ee}
\end{figure}

\subsection{Size and shape}
\label{sec_sz_shp}
Associated with the active open to spiral transition, the polymer undergoes significant change in its size and shape. In this section we clearly demonstrate these transformations with the help of ($i$)~the end- to- end separation, and ($ii$)~the radius of gyration tensor.   

\begin{figure}[t]
	\centering
		\includegraphics[width=8.6cm]{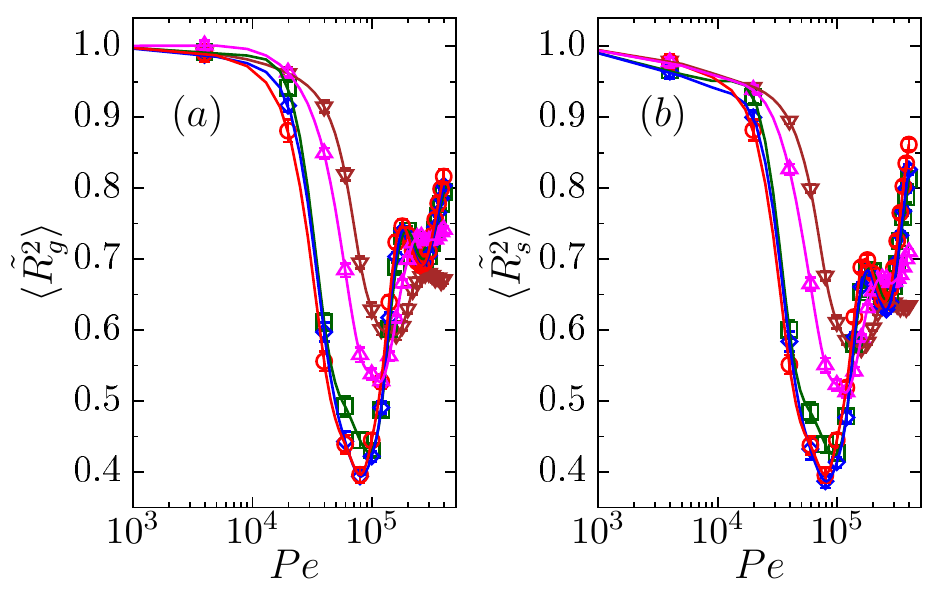}
		\caption{(color online) Radius of Gyration. Fig. $(a),(b)$ shows the variation of size $\tilde{\langle R_{g}^{2}\rangle}$ and shape $\tilde{\langle R_{s}^{2}\rangle}$ with standard error as a function of $Pe$ at different $\omega_{on}/\omega_{0}=0.5$ (brown $\triangledown$), $1$ (pink $\triangle$) $5$ (green $\Box$), $10$ (blue $\Diamond$), $20$ (red $\Circle$) respectively. 
		}
		\label{fig_rgrs}
\end{figure}

\subsubsection{End- to- end separation}
\label{sec_ee}
In Fig.~\ref{fig_ee}($a$) we show the probability distribution of the scaled end-  to- end separation $\tilde r_e = r_e/\la L \ra$ of the polymer for different $Pe$ at a fixed $\on/\w_0$, where $\la L \ra$ denotes the mean contour length. The distribution function $p(\tilde r_e)$ is normalized to $\int_0^1 d\tilde r_e\,2\pi \tilde r_e\, p(\tilde r_e)=1$. At $Pe=0$, it shows a single maximum at $\tilde r \approx 0.8$ corresponding to rigid- rod like configurations. 
This points to a relatively large effective bending rigidity of the filament~\cite{Gupta2019}. Note that $Pe = 0$ does not imply an equilibrium passive polymer,  because of the active attachment- detachment of the MPs with $\on/\w_0\neq 0$.
With increasing $Pe$, the distribution changes qualitatively. At $Pe=0.2 \times 10^5$, 
a new maximum appears near  $\tilde r \approx 0.15$. This bimodality corresponds to coexistence of rod-like shapes with folded polymers, a behaviour that appears even before the chain starts to form spirals. At further higher activity, $Pe \geq 0.4 \times 10^5$, as the probability of spiral- state increases, the small $\tilde r_e$ maximum shifts to smaller values, and their corresponding probability increases up to $Pe=1.19\times 10^5$. At even higher $Pe$, the spiral state starts to become less stable, as has been discussed in Sec.~\ref{sec_pt}. Associated with that, the height of the small $\tilde r_e$ maximum in $p(\tilde r_e)$ decreases. This non-monotonic behaviour is clearly observable in Fig.~\ref{fig_ee}($a$). The peak at small $\tilde r_e$ increases with increasing activity in the range of $Pe\times 10^{-5}=0.2$ to $1$. At higher activity,   $Pe\times 10^{-5}=0.99,\, 1.98,\, 3.57$, this peak- height decreases.     

In  Fig.~\ref{fig_ee}($b$), we show the non-monotonic variation of the second moment of the end- to- end distribution $\la \tilde r_e^2 \ra$ with $Pe$, for different $\on/\w_0$. 
For all $\on/\w_0$, $\la \tilde r_e^2 \ra$ initially decreases from the value at $Pe = 0$ as the polymer starts folding and getting into predominantly spiral states. As $Pe$ is increased further,  $\la \tilde r_e^2 \ra$  starts increasing since the stability of spiral states decrease. At small  $\on/\w_0$, $\la \tilde r_e^2 \ra$ shows eventual saturation with $Pe$. However, for larger $\on/\w_0$, the curve shows a further non monotonic behavior with an asymptotic increase in $\la \tilde r_e^2 \ra$  at higher values of $Pe > 2.58 \times 10^5$.
Note that a non-monotonic variation of  $\la \tilde r_e^2 \ra$ with $Pe$ was observed earlier in polymers in active bath~\cite{Eisenstecken2016}. The main difference   of that result with our model is, for $\on/\w_0 \geq 5$ we find two minima in the $\la \tilde r_e^2 \ra$ versus $Pe$ curve instead of the single minimum in Ref.~\citenum{Eisenstecken2016}, before the asymptotic increase. 
The size variation is associated with the effective persistence length of the filament~(see Appendix-\ref{sec_tt}).

\subsubsection{Radius of gyration tensor}
\label{sec_rg}
The size and shape of the polymer configurations can be extracted from analyzing the 
 radius of gyration matrix 
\bea
 S=\frac{1}{N}
\begin{pmatrix}
 \sum_{i} (x_{i}-x_{cm})^{2} &  \sum_{i} (x_{i}-x_{cm})(y_{i}-y_{cm})\\
 \sum_{i} (x_{i}-x_{cm})(y_{i}-y_{cm}) & \sum_{i} (y_{i}-y_{cm})^{2}
\end{pmatrix}\nn\\
\label{Rg matrix}
\eea
where $(x_i, y_i)$ denotes the position vector of the $i$-th bead, and $(x_{cm}, y_{cm})$ denotes the center of mass coordinate of the instantaneous polymer configuration. The two eigenvalues $\tilde \l_+$ and $\tilde \l_-$ of $S/\la L \ra^2$ describe the instantaneous configuration of the polymer as an elliptical shape, with $\tilde \l_+$ and $\tilde \l_-$ denoting the square of lengths along the semi-major and semi-minor axes whose orientations are determined by the eigenvectors. 
A measure of effective size of the polymer is given by $R^2_{g}=\tilde \l_+ + \tilde \l_-$. The difference between the eigenvalues denotes its shape $R_s^2 = \tilde \l_+ - \tilde \l_-$, with $R_s^2 =0$ for a symmetric circular shape.  
In Fig.~\ref{fig_rgrs} we show variations of the scaled size $\la \tilde R_g^2 \ra = \la R_g^2 \ra/ \la R_g^2 \ra_{Pe=0}$ and shape $\la \tilde R_s^2 \ra= \la R_s^2 \ra/\la R_s^2 \ra_{Pe=0}$ with $Pe$. 
As expected, the variation of $\la \tilde R_g^2 \ra$ follows the same non-monotonic variation as the other measure of size $\la \tilde r_e^2\ra$ shown in Fig.~\ref{fig_ee}. Remarkably, the shape of the polymer $\la \tilde R_s^2 \ra$ follows the same qualitative dependence on $Pe$ at all $\on/\w_0$ ratios. 
See Appendix-\ref{sec_distb} for probability distributions of $\tilde \l_+$, $\tilde \l_-$, $\tilde R_g^2$ and $\tilde R_s^2$. 

\subsection{Dynamics}
\label{sec_dyn}
Associated with the re- entrant phase transition, the conformational dynamics displays a  non-monotonic variation of the characteristic time- scales with MP activity. In this section, we study the two- time autocorrelation functions corresponding to the turning number, the radius of gyration, and the polymer shape as defined above. The overall orientation, described by the eigenvector corresponding to the larger eigenvalue of the radius of gyration tensor, does not involve conformational relaxation. As a result, its dynamics gets faster monotonically with increasing activity. 

\begin{figure}[t]
	\centering
	\includegraphics[width=8cm]{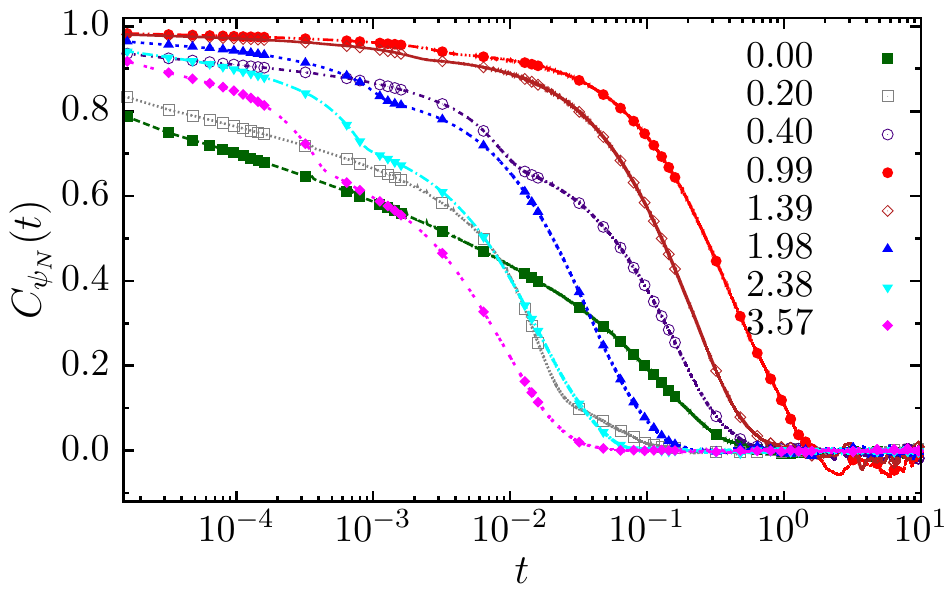}
	\caption{(color online) Two time autocorrelation function $C_{\psi_N}(t) = \la \psi_N(t) \psi_N(0)\ra/ \la \psi^2_N(0) \ra$ evaluated at different $Pe = \tilde{Pe}\times 10^5$ with $\tilde{Pe}$ shown in the figure- legend keeping the on- off ratio $\on/\w_0 = 1$  constant. Time $t$ is expressed in the unit of $\t$.
	}
	\label{fig:corr_op}
\end{figure}

\begin{figure*}[!htbp]
	\centering
	\includegraphics[width=18cm]{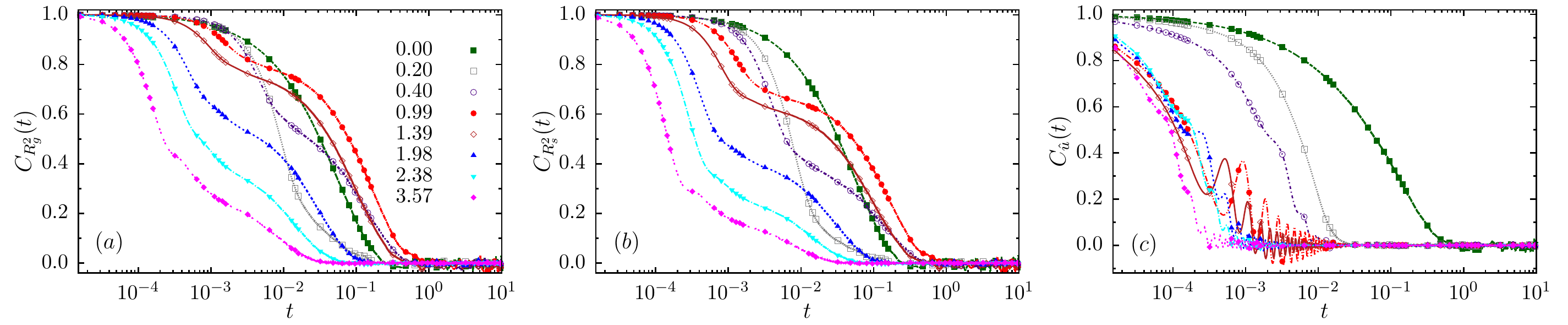} 
		\caption{(color online) Two- time correlation functions of $(a)$~$R_{g}^{2}$,  ($b$)~$R_{s}^{2}$, and ($c$)~eigen- vector $\uv$ at various $Pe=\tilde{Pe} \times 10^5$ with $\tilde{Pe}$-values indicated in the legend, keeping $\on/\w_0=1$ fixed. Time $t$ is expressed in the unit of $\t$.}
\label{fig_corr}		
\end{figure*}

\subsubsection{Dynamics of turning number} 
\label{sec_corr1}
In Fig.~\ref{fig:corr_op} we show the two- time autocorrelation function of the turning number,
$C_{\psi_N}(t) = \la \psi_N(t) \psi_N(0) \ra / \la \psi^2_N(0) \ra$ at different $Pe$ values keeping the ratio $\on/\w_0 = 1$ constant. In using this definition it is noted that $\la \psi_N(t) \ra = 0$ by symmetry, thus the fluctuation $\d \psi_N = \psi_N$. 
For $Pe \leq 0.2 \times 10^5$, the chain stays in the open state corresponding to the unimodal distribution  in $p(\psi_N)$ with the maximum at $\psi_N=0$. The stochastic relaxation within this state gives rise to the single- exponential decay observed in Fig.~\ref{fig:corr_op}.  
At the phase coexistence, 
{ a new mechanism corresponding to the switching between the open and spiral states can lead to a crossover of the correlation to a second exponential decay. In Fig.\ref{fig:corr_op} we observe such a double exponential for $Pe \geq 0.4 \times 10^5$. 
} 
The crossover between the two exponentials gets imperceptibly shallow between $Pe=10^5$ and $2\times 10^5$, as the overall faster dynamics due to larger $Pe$, makes the switching between states easier. 
As we increase $Pe$ further, the increasing number of turns of the polymer increases the distance between the open and spiral peaks in the space of  $\psi_N$. Switching between states becomes prohibitively expensive which makes the crossovers sharper again.

\subsubsection{Dynamics of size, shape, and orientation}
The dynamics of the size, shape, and overall orientation of the polymer can be determined by analyzing the time- series of the eigenvalues of the radius of gyration tensor, and the eigen- vector  $\uv$ corresponding to the larger eigenvalue $\l_+$. We use the correlation functions 
$C_{R_g^2}(t) = \la \d R_g^2(t) \d R_g^2(0) \ra/ \la \d R_g^4 \ra$,
$C_{R_s^2}(t) = \la \d R_s^2(t) \d R_s^2(0) \ra/ \la \d R_s^4 \ra$,
and
$C_{\uv}(t) = \la \uv(t) \cdot \uv(0) \ra$.
The fluctuations $\d R_{g,s}^2(t) = R_{g,s}^2(t) - \la R_{g,s}^2(t) \ra$. 
The correlation functions are plotted in Fig.~\ref{fig_corr}. The size and shape correlations 
display double- exponential decay at $Pe \geq 0.4 \times 10^5$, as in the turning number correlation function in Fig.~\ref{fig:corr_op}. This is because of the close relation between the size, shape and the turning number, all of which depend on the polymer conformation. 

However, the dynamics of the overall orientation of the polymer captured by $\uv$, is not related to internal structural relaxation. Thus it shows single exponential decay of the correlation, describing an orientational diffusion at $Pe \leq 0.2\times 10^5$. Once the spirals are formed  they start to rotate under the active drive. As a result, the orientation $\uv$ also rotates. This is captured by the oscillations in $C_{\uv}(t)$ at $Pe \gtrsim 0.4\times 10^5$. As can be easily seen from Fig.~\ref{fig_corr}($c$), the frequency of rotation increases and the amplitude of oscillation in $C_{\uv}(t)$ decreases with increasing $Pe$.

\subsubsection{Time scales}
\label{sec_ts}
\begin{figure}[!t]
	\centering
	\includegraphics[width=9cm]{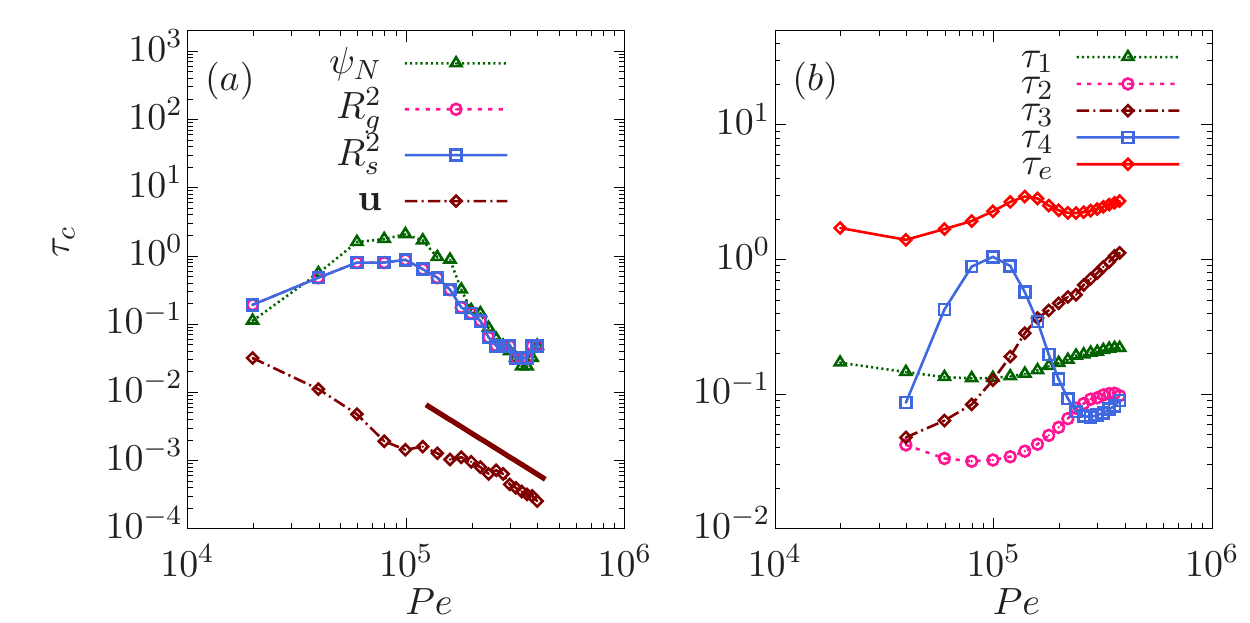} 
	\caption{(color online)  ($a$)~Variation of correlation time $\t_c$ (in the unit of $\t$) with $Pe$ at $\on/\w_0=1$,  obtained for $\psi_N$, $R_g^2$, $R_s^2$, and $\uv$. The brown solid line denotes the scaling form $1/Pe^2$. ($b$)~Time scales calculated at $\on/\w_0=1$ using the expressions from the approximate non-conserved dynamics Eq.\eqref{eq_nc}.  
	}
	\label{fig_corrt}
\end{figure}

The correlation time $\t_c$ is the time scale at which the autocorrelation function touches zero for the first time. In Fig.~\ref{fig_corrt}($a$) we show the dependence of $\t_c$ on $Pe$ keeping $\on/\w_0=1$. $\t_c$ corresponding to the orientational correlation function $C_{\uv}(t)$ decreases monotonically with increasing $Pe$. This can be understood by noticing that the overall orientational dynamics does not involve internal conformational relaxation of the polymer. It is thus controlled by the active time scale $D/v_0^2 \sim 1/Pe^2$, and decreases monotonically with increasing $Pe$~(Fig.~\ref{fig_corrt}($a$)\,).

On the other hand, the value of $\t_c$ corresponding to $C_{\psi_N}(t)$, $C_{R_g^2}(t)$ and $C_{R_s^2}(t)$ is controlled  by two competing effects. The enhanced activity at higher $Pe$  is expected to make the dynamics faster. On the other hand, as the system undergoes phase transition, the slow switching between states can slowdown the overall dynamics. This competition leads to a non-monotonic variation of $\t_c$ with a maximum 
reached at $Pe = 10^5$~(Fig.~\ref{fig_corrt}($a$)\,). The maximum in $\t_c$ is associated with the dominance of spirals in the dynamics. The correlation time $\t_c$ for a smoothened chain of $r_0/\s=0.75$ shows a similar non-monotonic variation~(Appendix-\ref{sec_tc_smooth}), however, with smaller $\t_c$ values than the chain with $r_0/\s=1.0$, due to a reduced sliding friction.

At this point, it is instructive to focus on $\t_c$ corresponding to $C_{\psi_N}(t)$. Note that at $Pe=10^5$, where the maximum of $\t_c = 1.3 \times 10^5\,\t$ is observed~(Fig.~\ref{fig_corrt}($a$)\,), the simulation results for the mean dwell times at the open and spiral states are $\t_o = 1.3 \times 10^4\,\t$ and $\t_s = 2.5 \times 10^4\,\t$, respectively. 
Using an assumption of a dichotomous Markov process, they lead to an estimate of the correlation time~\cite{Gardiner1983} $\t_e=\t_o \t_s/(\t_o + \t_s) \approx 10^4\,\t < \t_c=1.3 \times 10^5\,\t$. Such a difference is not unexpected as the actual dynamics is not really a dichotomous process, and involves other mechanisms, e.g, a gradual transition between the open and spiral states.   

In the following we attempt to obtain estimates of $\t_o$ and $\t_s$ using a relaxation dynamics corresponding to the effective free energy $\cf(\psi_N)$ in Eq.\eqref{eq_fe}.
For notational simplicity, we replace $\psi_N$ by $\psi$ in the rest of this section. 
The non-conserved dynamics is given by~\cite{Chaikin1995} 
\bea 
\p \psi/\p t = - M \left[ \p \cf/\p \psi \right] + \sqrt{2 \kb T_e M}\, \L(t),
\label{eq_nc}
\eea 
where $T_e$ plays the role of an effective temperature, $M$ a mobility and $\L(t)$ is a univariate  and uncorrelated Gaussian random noise.  
The triple- minima of the {\em free energy} are at 
$\psi = 0$ and 
$\psi_m = \pm ({u_4}/{3 u_6})^{1/2} \left[ 1+\left( 1- {3 u_2 u_6}/{2 u_4^2} \right)^{1/2} \right]^{1/2}$, while the double maxima are at
$\psi_M = \pm ({u_4}/{3 u_6})^{1/2} \left[ 1-\left( 1- {3 u_2 u_6}/{2 u_4^2} \right)^{1/2} \right]^{1/2}$.
Disregarding the mobility $M$ in the absence of an independent measure, the relaxation time scales at the minima of $\cf$, are given by $[\p^2 \cf/\p \psi^2]^{-1}_{\psi=0,\psi_m}$. 
The relaxation around $\psi=0$ leads to the inverse time-scale $\t_1^{-1} \sim \w_1 = u_2$, 
and that around  $\psi=\psi_m$ gives 
$\t_2^{-1} \sim \w_2= u_2 - (4 u_4^2/u_6)  \left[ 1+\left( 1- {3 u_2 u_6}/{2 u_4^2} \right)^{1/2} \right] + (30 u_4^2/9 u_6) \left[ 1+\left( 1- {3 u_2 u_6}/{2 u_4^2} \right)^{1/2} \right]^2$. 
The expressions for $\t_1$ and $\t_2$ at $\on/\w_0=1$ are plotted in Fig.~\ref{fig_corrt}($b$). 
Further, we calculate the Kramer's escape times~\cite{Gardiner1983} for barrier crossing: $\t_3$  from $\psi=0$, and $\t_4$ from $\psi=\psi_m$. These  are 
$\t_3 \sim (\w_1 |\w_M|)^{-1} \exp[\cf(\psi_M) - \cf(0)]$, 
and $\t_4 \sim (\w_2 |\w_M|)^{-1}\exp[\cf(\psi_M) - \cf(\psi_m)]$, 
where 
$\w_M = u_2 - (4 u_4^2/u_6)  \left[ 1-\left( 1- {3 u_2 u_6}/{2 u_4^2} \right)^{1/2} \right] + (30 u_4^2/9 u_6) \left[ 1-\left( 1- {3 u_2 u_6}/{2 u_4^2} \right)^{1/2} \right]^2$~(see Fig.~\ref{fig_corrt}($b$)\,). It is interesting to note that, among these time scales, only $\t_4$, {the time- scale determining the rate of exiting the spiral state,} has a non-monotonic variation with $Pe$, { and dominates the overall behavior}.  
The above analysis allows us to express the two effective dwell times as 
$\t_o = (\t_1 + \t_3)$ and $\t_s = (\t_2 + \t_4)$. 
The estimate $\t_e=\t_o \t_s/(\t_o + \t_s)$ is plotted in Fig.~\ref{fig_corrt}($b$) with a multiplicative shift by $10$ for better visibility. This shows a non-monotonic variation, with a small maximum at  an intermediate $Pe$, a behavior that is qualitatively similar to the dependence of correlation times corresponding to $\psi_N$, $R_g^2$ and $R_s^2$ with $Pe$~(Fig.~\ref{fig_corrt}($a$)\,). 

The main caveat to the above analysis is Eq.\eqref{eq_nc} obeys the equilibrium fluctuation- dissipation relation, and is not strictly valid as a description for active systems. Further, even within an effective equilibrium interpretation, the Kramer's theory of barrier crossing is subject to modification when interpreted for transition rates between multiple minima of a free energy profile. 

\section{Discussion}
\label{sec_dis}
We considered a detailed model of motility assay consisting of an extensible semiflexible filament driven by motor proteins (MP) immobilized on a substrate. The numerical simulations showed a reentrant first order transition from open chain to spirals with changing activity. This transition is characterized by the presence of metastable maxima in the probability distribution of turning number.
We obtained the phase- diagram in the $Pe\,- \on/\w_0$ plane, which clearly brings out the importance of attachment- detachment kinematics of the MPs. 
At a constant $\on/\w_0$, the polymer shows reentrance transition from open chain to spiral to open chain with increasing $Pe$. With lowering of $\on/\w_0$, the phase boundary shifts  progressively to higher $Pe$, following a hyperbolic relation 
derived from a local torque balance. 

The reentrant transition is associated with non-monotonic variations of the polymer size, shape
and fluctuations in turning number $\la \psi_N^2\ra$ with $Pe$.  The data collapse of the $\la \psi_N^2\ra$ versus $Pe$ curves at different $\on/\w_0$ led to a scaling relation, which could  approximately be captured by the torque balance argument that describes the phase boundary. The coexistence of open chain and spirals is preceded by a coexistence of open and folded chains captured by the bimodality in the distribution of end- to- end separation.     

Our detailed analysis of the dynamics showed a double- exponential decay in the autocorrelation function of size, shape and turning number.  
The corresponding correlation times showed a non-monotonic variation with $Pe$, { with a maximum due to the dominance of spirals}. We developed an approximate description of the correlation time in terms of a dichotomous process between the open and spiral states. Using an effective free energy description of the phase transition and non-conserved relaxation dynamics, we obtained expressions for the dwell times in the two states, giving an estimate of the correlation time. This showed a non-monotonic variation with $Pe$, albeit  with relatively small variations. The two-time autocorrelation of the polymer orientation, on the other hand, showed a single exponential decay, with characteristic oscillations associated with the rotation of spirals. The orientational dynamics does not depend on the conformational relaxation, and the  corresponding correlation time decreases with activity as $1/Pe^2$.

Our detailed modeling of MPs allowed us to explicitly identify dependence of the polymer properties on both the active velocity of MPs $v_0$, and the attachment- detachment kinematics fixed by the ratio $\on/\w_0$. Together, they characterize the MP activity and depend on the ambient ATP concentration. 
Our predictions are amenable to direct experimental verifications in {\em in vitro} motility assays. 
For example, we can estimate the correlation time for turning number and polymer extension of a filament driven by motor proteins. The viscosity in the cell is around 100 times that of water $\eta_w = 0.001\,$pN-s/$\mu{\rm m}^2$~~\cite{Howard2005}.
Assuming a similar viscosity in the motility assay, one gets $\eta=100\,\eta_w = 0.1\,$pN-s/$\mu{\rm m}^2$. The corresponding viscous damping over a bond-length $\s$ is $\g=3\pi\eta\s$. The activity of MPs can be changed by changing the ambient ATP concentration. For example, for kinesins, the active velocity $v_0$ varies from $0.01\,\mu$m/s to $1\,\mu$m/s, as the ATP concentration is increased from $1\,\mu$M to $1\,$mM~\cite{Schnitzer2000}. This corresponds to $Pe=\g v_0 L^2/\kb T \s = 3\pi \eta v_0 L^2/\kb T$. At room temperature $\kb T = 4.2 \times 10^{-3}$pN-$\mu$m.  A filament of length $10\,\mu$m experiences $Pe\approx 2 \times 10^4$.  Using the unit of time $\t=\g L^3 /4 \s \kb T = 3 \pi \eta L^3/4 \kb T \approx 15.6$~hours, the estimated correlation time for turning number, radius of gyration and end-to-end separation of the filament $\sim 0.1\,\t$ translates to about 1.5 hours. Our qualitative predictions for transitions between open chain to spiral, and the non-monotonic variations of the polymer size and shape with changing $v_0$ can be tested by controlling  ATP concentration in the motility assays.

\section*{Conflicts of interest}
There are no conflicts to declare.

\section*{Acknowledgements}
The computations were supported in part by SAMKHYA, the high performance computing facility at Institute of Physics, Bhubaneswar. D.C. thanks SERB, India for financial support through grant numbers MTR/2019/000750 and EMR/2016/001454.  DC and AC thank International Centre for Theoretical Sciences (ICTS) for support during a visit for participating in the program - Thirsting for Theoretical Biology (Code: ICTS/ttb2019/06).

\appendix

\begin{figure}[!h]
	\centering
	\includegraphics[width=8cm]{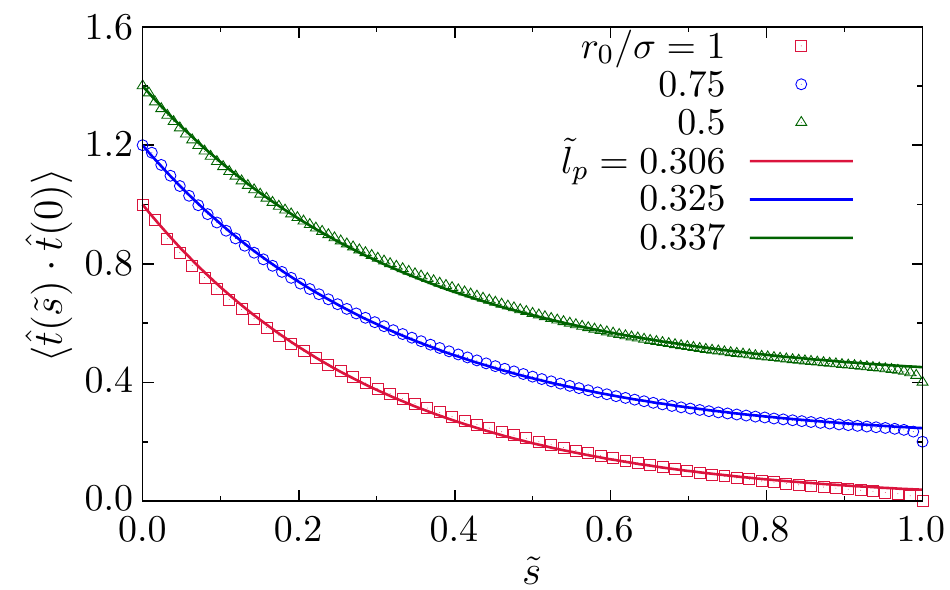}
	\caption{(color online) Increase in equilibrium persistence length due to reduction of bond length $r_0$ with respect to the WCA length scale $\s$. Tangent-tangent correlation functions $\la \hat t (\tilde s) \cdot \hat t (0)\ra$ as a function of $\tilde s = s/L$ are shown at three values of $r_0/\s = 1,\,0.75,\, 0.5$. The lines denote the exponential decay of correlation as $\exp(-\tilde s/\tilde l_p)$, with corresponding persistence length $\tilde l_p$ denoted in the figure legend. Plots for $r_0/\s = 0.75,\, 0.5$ are shifted upwards by $0.2$ and $0.4$ for better visibility. 
 }
\label{eq_lp}
\end{figure}

\section{Equilibrium persistence}
\label{sec_lp}
{
The presence of WCA repulsion between non-bonded beads changes the equilibrium properties of the chain with respect to an ideal semiflexible polymer. At short length scale it increases the effective persistence length. This can be seen from Fig.~\ref{eq_lp}, where we plotted the tangent-tangent correlation $\la \hat t (\tilde s) \cdot \hat t (0)\ra$ with $\tilde s = s/L$ denoting relative contour-wise separation between bonds. Smoothening of the potential profile along the chain, reducing bond length $r_0$ with respect to the WCA size $\s$, leads to enhanced repulsion between neighboring bonds. This adds to the energy cost to transverse fluctuations, thereby increasing the effective persistence length $\tilde l_p$ defined as $\la \hat t (\tilde s) \cdot \hat t (0)\ra \approx \exp(-\tilde s/\tilde l_p)$. The increase in $\tilde l_p$ with $r_0/\sigma$ is shown in Fig.~\ref{eq_lp}.  In fact, for longer chains, at   large contour separations with respect to the persistence length, the effect of self-avoidance dominates over bending rigidity. It leads to the Flory scaling $\la r^2(s) \ra \sim s^{2\nu}$, which corresponds to a power-law decay in the correlation $\la \hat t (s) \cdot \hat t (0)\ra \sim s^{-(2-2\nu)}$ at long contour separations. In an intermediate $\tilde s$, the correlation function crosses over from exponential to power-law decay.  
}

\begin{figure}[t!]
	\centering
	\includegraphics[width=8.5cm]{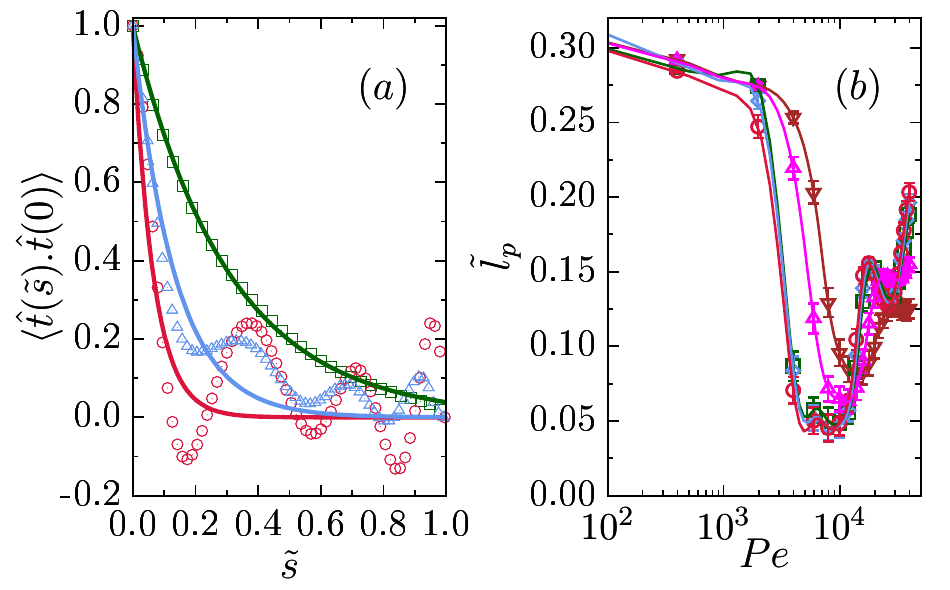} 
	\caption{(color online) Persistence length. $(a)$~Tangent-tangent correlation function for $Pe \times 10^{-5}=0(\Box)$, $0.99(\Circle)$, $1.98(\triangle)$ and $\omega_{on}/\omega_{0}=1$. The points denote the simulation results, and the solid lines represent the fitting functions $\exp(-\tilde s/\tilde{l}_{p})$. $(b)$~Variation of the effective persistence length $\tilde{l}_{p}$ with $Pe$ at $\on/\w_0=0.5\,(\triangledown)$, $1\,(\triangle)$, $5\,(\Box)$, $10\,(\Diamond)$, $20\,(\Circle)$.  The lines through data are guides to eye. At equilibrium, the chain has persistence length $\tilde l_p \approx 0.3$, close to the values at $Pe=0$. }
	\label{fig_tt}
\end{figure}

\begin{figure}[b!]
	\centering
	\includegraphics[width=8.6cm]{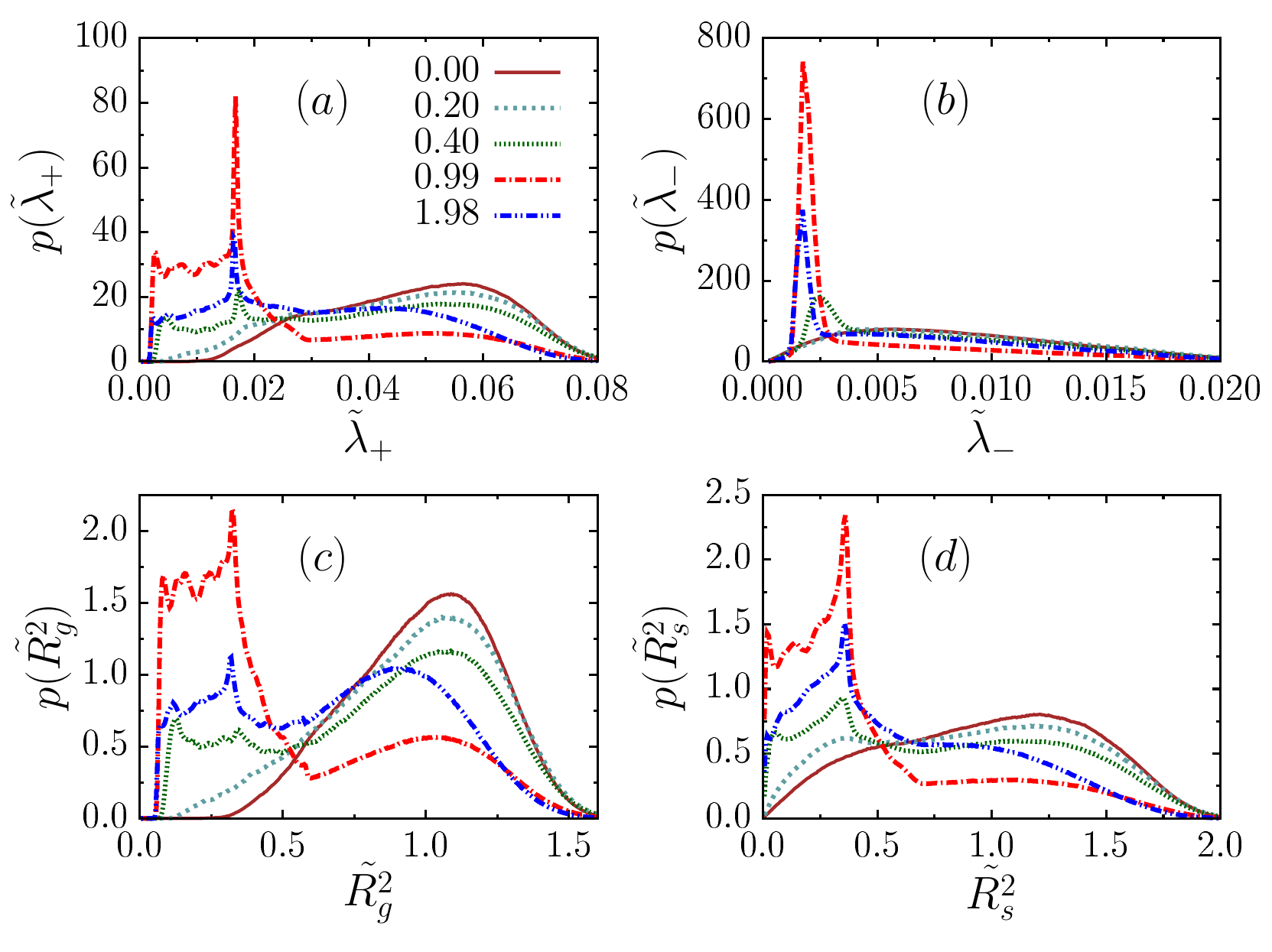}
	\caption{(color online) Probability distributions of the eignevalues ($a$)~$\tilde \l_+$, and ($b$)~$\tilde \l_-$, ($c$)~size $\tilde R_g^2$, and ($d$)~shape $\tilde R_s^2$ are shown. All the distribution functions are obtained at $\on/\w_0=1$, and the different graphs in ($a$)--($d$) correspond to the $Pe = {\tilde Pe} \times 10^5$ with ${\tilde Pe}$- values indicated in the legend of $(a)$. }
	\label{fig_plam}
\end{figure}

\section{Effective persistence length}
\label{sec_tt}
The effective persistence length can be quantified in terms of the correlation function between local tangents $\la \tv(s) \cdot \tv(0) \ra$ at contour positions separated by $s$. For the worm- like- chain the correlation shows a single- exponential decay $\la \tv(s) \cdot \tv(0) \ra = \exp(-s/\l)$  defining the persistence length $\l$. The semiflexible polymer under
the motility assay drive shows non- trivial tangent correlations~(Fig.~\ref{fig_tt}($a$)\,). The oscillations in the decaying correlation at higher $Pe$ is associated with the formation of the spiral configurations. However, the initial decay in correlation can be fitted to a single exponential form $\exp(-s/l_p)$ to capture the effective persistence length $l_p$. In Fig.~\ref{fig_tt}($a$) the contour lengths $s$ are expressed as  $\tilde s = s/\la L \ra$, where $\la L \ra$ is the mean chain- length. The scaled effective persistence lengths $\tilde l_p = l_p/ \la L \ra$ are plotted as a function of $Pe$, at fixed $\on/\w_0$ ratios in  Fig.~\ref{fig_tt}($b$). The variation of  $\tilde l_p$ shows non-monotonic change with $Pe$, and follows the variation of the mean squared end- to- end separation $\la \tilde r_e^2 \ra$ plotted in Fig.~\ref{fig_ee}($b$).

\section{Radius of gyration: Probability distributions}
\label{sec_distb}
Here we show the probability distributions of the eigenvalues of the radius of gyration matrix, $p(\l_\pm)$. In Fig.~\ref{fig_plam}($a$), ($b$) we show these distribution functions evaluated for various $Pe$ and a fixed turnover $\on/\w_0=1$. Clearly, at the onset of instability towards formation of spirals both the distributions $p(\tilde \l_\pm)$ start to show emergence of a very sharp delta- function like peak. This corresponds to a typical size and shape of the configurations forming spiral. This feature is further quantified in the distribution functions of the relative size and shape variables $\tilde R_g^2$ and $\tilde R_s^2$ shown in  Fig.~\ref{fig_plam}($c$), ($d$).

\begin{figure}[!t]
	\centering
	\includegraphics[width=8.6cm]{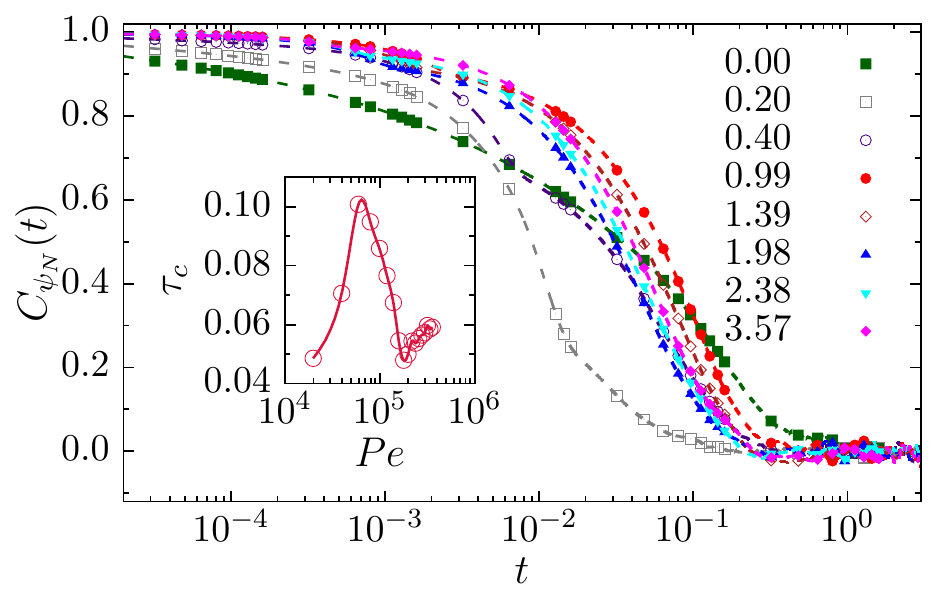} 
		\caption{(color online) Two time correlation functions of turning number $C_{\psi_N}(t)$ for a smoothened polymer with $r_0/\s=0.75$ calculated at $\on/\w_0=1$ and various activity $Pe=\tilde Pe \times 10^5$ with $\tilde Pe$-values denoted in the figure legend.  The inset shows a  non-monotonic variation of the corresponding correlation times with $Pe$.}
	\label{corr_75}
\end{figure}

{
\section{Correlation time in smoothened polymer}
\label{sec_tc_smooth}
Here we compute the correlation function of turning number $C_{\psi_N}(t)=\la \psi_N(t) \psi_N(0)\ra/\la \psi_N^2(0) \ra$ for the smoothened chain with $r_0/\s=0.75$~(Fig.\ref{corr_75}) at different values of $Pe$ keeping the attachment- detachment ratio $\on/\w_0=1$ fixed. To keep the chain length unchanged with respect to the chain with $r_0/\s=1.0$, we use $N=85$ beads. The correlation times $\t_c$ are determined by identifying where  $C_{\psi_N}(t)$ touches zero. The plot of correlation time in the inset of Fig.\ref{corr_75} shows a non-monotonic variation similar to Fig.\ref{fig_corrt}($a$), while the actual values of $\t_c$ remains smaller than the chain with $r_0/\s=1.0$. 
}
\bibliographystyle{prsty}

\end{document}